\documentstyle[epsfig,epsf,twocolumn,aps]{revtex}
\tightenlines

\newcommand{\be}{\begin{equation}}
\newcommand{\ee}{\end{equation}}
\begin{document}
\twocolumn[\hsize\textwidth\columnwidth\hsize\csname @twocolumnfalse\endcsname
%\thispagestyle{empty}
%\setcounter{page}{0}
%\begin{flushright}
%\end{flushright}

%\vspace{3mm}

\title{Fermi Surfaces of Diborides: MgB$_2$ and ZrB$_2$
       }

\author{H. Rosner$^{\dag}$, J. M. An$^{\dag}$, W. E. Pickett$^{\dag}$
      and S.-L. Drechsler$^{\ddag}$} 

\address{$^{\dag}$Department of Physics, University of 
                     California, Davis CA 95616}
\address{$^{\ddag}$Institut f\"ur Festk\"orper- und Werkstofforschung,
     P. O. Box 270116, D-01171 Dresden, Germany}

\date{\today}
\maketitle
\tightenlines
\begin{abstract}
We provide a comparison of accurate full potential band calculations
of the Fermi surfaces areas and masses of MgB$_2$ and ZrB$_2$ with the
de Haas-van Alphen date of Yelland {\it et al.} and Tanaka {\it et
al.} respectively.  The discrepancies in areas in MgB$_2$ can be
removed by a shift of $\sigma$ bands downward with respect to $\pi$
bands by 0.24 eV.  Comparison of effective masses lead to orbit
averaged electron-phonon coupling constants $\lambda_{\sigma}$=1.3
(both orbits), $\lambda_{\pi}$=0.5.  The required band shifts, which
we interpret as an exchange attraction for $\sigma$ states beyond
local density band theory, reduces the number of holes from 0.15 to
0.11 holes per cell.  This makes the occurrence of superconductivity
in MgB$_2$ a somewhat closer call than previously recognized, and
increases the likelihood that additional holes can lead to an
increased T$_c$.
\end{abstract}
%\centerline{DRAFT---Not for Distribution}
\vskip 1cm
\footnoterule
%$^{\ddag}$ \parbox[t]{6in}{
%{\small E-mail: ~~pickett@physics.ucdavis.edu} } 
%\pacs{71.20.Lp,75.10.Lp,75.30.-m,75.50.Gg}
\newpage

% for twocolumn activate the line below...
]
%\begin{multicols}{2}
\section{Introduction}
The discovery of superconductivity in MgB$_2$ near 40 K by Akimitsu's
group\cite{akimitsu} and the subsequent intense experimental
investigation of its properties and theoretical exposition of the
underlying causes, has made it clear that MgB$_2$ is the first member
of a new class of superconductors.  Although intermetallic like the
best superconductors before 1986, it does not have $d$ electrons, it
does not have a high density of states at the Fermi level E$_F$, and
it is strongly anisotropic in its crystal and electronic 
structures.\cite{jan,kortus,kong,bohnen,yildirim,shulga,ravindran}
In another important way, it is distinct from the other intermetallic
superconductors: it derives its high superconducting critical
temperature T$_c$ from extremely strong coupling from only a small
fraction of the phonons to specifically a limited part of the Fermi
surface (FS).\cite{jan,kong,bohnen,yildirim,liu,choi}  
Other diborides, at least so far, are disappointing
with regard to their superconductivity -- a study of TaB$_2$ reveals
that the presence of Ta $d$ electrons in the valence band region
results in a completely different electronic structure,\cite{tab2}
especially the states at the Fermi level, and accounts for its lack of
bulk superconductivity.

Although the properties of MgB$_2$ appear to be described
consistently, and reasonably accurately, by a Fermi liquid picture
based on the band structure calculated in the local density
approximation (LDA), there have been few opportunities to make
detailed quantitative comparison with experimental data.  Therefore
there has been a broad range of alternative suggestions. 
Imada has suggested strong interband Coulomb
exchange processes\cite{imada}. Furukawa has raised the possible
importance of half-filled $p_z$ bands\cite{furukawa}. Hirsch and
Marsiglio\cite{hirsch} suggest that hole-undressing by Coulomb
interaction in nearly filled bands is responsible.  Referring to
optical data on oriented films, Marsiglio\cite{frank} has suggested
that coupling via a high energy electronic mode is plausible.
Baskaran\cite{baskaran} has revived the Pauling
resonating-valence-bond picture of benzene as having possible
application to the graphene layers of boron in MgB$_2$.  
Nonadiabatic processes strongly affecting the occurrence of superconductivity
have been put forward, by Alexandrov based on penetration depth data
\cite{alexandrov} and by Cappelluti {\it et al.} on the basis of several
experimental results that are not readily understandable in terms of
conventional (isotropic) Eliashberg theory.\cite{cappelluti}

Although the
crystal structure of MgB$_2$ is quite simple, it is strongly layered
and electronic characteristics are predicted to be strongly
anisotropic.  So far few single crystals have been obtained, so single
crystal optical conductivity data is not available.  In addition, the
strong ``anisotropies'' mentioned above necessitate a two-band
model\cite{shulga,liu,golubov,junod} or a fully anisotropic
treatment\cite{choi} to account properly for the data, and such
interpretations are only recently being carried out.  Photoemission
data on single crystals allow direct comparison, and the agreement
between band theory and angle-resolved photoemission spectroscopy is
good\cite{arpes,servedio} down to the scale of a few tenths of an eV.

The recently reported observation of de Haas-van Alphen (dHvA)
oscillations on single crystals by Yelland {\it et al.}\cite{dhva} is
a crucial development that provides the opportunity for detailed
evaluation of LDA predictions.  These measurements detected three
frequencies (extremal Fermi surface areas F) and information about the
orbit-averaged electron-phonon effective mass $m^*$ and scattering
time $\tau$. Comparing to reports of Elgazzar {\it et
al.}\cite{elgazzar}, who used the augmented spherical wave (ASW) method
within the spherical potential approximation, Yelland {\it et al.}
concluded that discrepancies with band theory are 40-80\% in the FS
areas.  This discrepancy seems large enough to suggest the occurrence
of important correlation effects beyond LDA.  If this is true, our
current understanding of the properties of MgB$_2$ might need
revision.  Mazin and Kortus have presented orbital areas\cite{mazin}
that are close enough to the observed values to give confidence in
the band picture.

In this paper we provide a comparison of careful
calculations of extremal areas and band masses for MgB$_2$.  To provide
comparison and contrast in a related diboride,
we provide similar information for the isostructural refractory diboride
ZrB$_2$, which has recently been reported\cite{gasparov}
to superconduct at T$_c$ = 5.5 K whereas earlier searches observed
no superconductivity.\cite{noscy} 
The extensive dHvA data available for ZrB$_2$ makes this a particularly
useful system to study.  We find, with one possible exception that we
discuss, that 
LDA predictions seem to provide an excellent  description for ZrB$_2$. 
For MgB$_2$ there is some
disagreement in FS areas that can be accounted for by a 
shift of $\pi$ (B $p_z$) bands with respect to 
$\sigma$ (B $sp_xp_y$) bands
by 240 meV and readjustment of the ``Fermi energies'' of each of these 
bands by $\sim \pm$ 120 meV.  This result is roughly 
consistent with that of Mazin and
Kortus.\cite{mazin}  Such shifts will lead to quantitative, but most likely not
qualitative, corrections in the extent explanation of MgB$_2$ 
superconductivity.

\section{Calculational Methods}

The precision and consistency of the calculations using different
methods is a concern, in light of the apparent discrepancy between the
band structure results of Elgazzar {\it et al.} and the data of
Yelland {\it et al.} Therefore we have applied two full
potential, all-electron methods of calculation that have produced
equivalent results for several other systems.\cite{mgcni3,comment} One
method is the full potential linearized augmented plane wave (FLAPW)
method\cite{djsbook} as implemented in the WIEN97 code.\cite{wien} The
other method is full potential local orbital code (FLPO) \cite{fplo}
based on local orbitals optimized to minimize the electronic energy.

There are two computational details that require attention, especially
for MgB$_2$, in order to obtain the precise predictions of band
theory.  Both are related to the existence of small volume Fermi
surfaces [the smallest contains $\sim$3\% of the Brillouin zone (BZ)
volume], for which small shifts of the band edge make appreciable
differences.  The first issue is that the non-spherical nature of the
charge density (related to the distinctly different contributions to
the density from $\sigma$ and $\pi$ states) and potential is
important.  Comparing the FPLO method with basis functions optimized
as usual (to minimize the energy) with the FLAPW results revealed that
the $k_z$ dispersion of the $\sigma$ band along $\Gamma$-A was
slightly different for the two methods. The reason is that the
extension parameter $x_0^{n,l}$ for the FPLO basis functions is
optimized only with respect to the main quantum number $n$ and the
angular momentum $l$ at a given site. Increasing the flexibility of
the FPLO basis resulted in agreement with the FLAPW result.  On the
other hand, the standard FPLO-basis set resulted in agreement with the
WIEN97 code for ZrB$_2$, suggesting that the charge density for the
latter compound is less anisotropic than in MgB$_2$. This implies as
well that the discrepancy of the results of Elgazzar {\it et
al.}\cite{elgazzar} is due to the spherical approximation of the
potential they used.

The second item is the dense $k$ point sampling that is necessary to
obtain the fraction of $\sigma$ band holes accurately, and hence the
charge density and potential, a point noted by Mazin and Kortus.\cite{mazin}
Increasing the number of inequivalent,
equally spaced points in the irreducible BZ from 2000 (which already
would be considered to be a fine mesh) to $\sim$16600 results in
changes in area of the smallest orbits (see Table I) by 2-3\%.
Because of such sensitivities, we quote areas and masses only to 2+
significant digits.  Unlike non-sphericity and k point sampling, the
choice of exchange-correlation potential makes no physical difference.  Using
the LAPW code, we checked the eigenvalues of the unoccupied $\sigma$
states at $\Gamma$ and A.  Measured relative to the Fermi level, these
differed by no more that 2.5 meV between the LDA of Perdew and
Wang\cite{perdewwang} and the generalized gradient approximation
of Perdew, Burke, and
Ernzerhof.\cite{gga}

\section{Results and Comparison with Experiment}
\subsection{ZrB$_2$}
As in MgB$_2$ (but for a different reason) it is not easy to prepare
single crystals of ZrB$_2$ due to its melting temperature above 3300 K.
However, single crystal studies have been reported.
dHvA data have been provided and analyzed for ZrB$_2$ by Tanaka and
Ishikawa,\cite{Zr_tanaka,Zr_ishizawa} and the data 
are similar to those reported for
the isostructural and isovalent sister compounds 
TiB$_2$\cite{Zr_ishizawa,Ti_tanaka}
and HfB$_2$.\cite{Hf_tanaka}

The band structure has been presented previously, based on a variety
of computational methods,\cite{ihara,harmon,anishchik,swit,shein} each
involving simplifications that we avoid.  Our calculated band
structure and density of states (DOS), calculated for $a$ =3.170 \AA,
$c$ =3.532 \AA, are similar to those presented earlier, so we do not
present them here.  The Fermi energy lies in a ``pseudogap,'' which is
due to relatively large velocities in the region of E$_F$ rather than
any semimetallic overlapping of bands (where there would be small FSs
and low velocities).
% The main part of the occupied B
%\sigma$ density of states (DOS) lies in the -6 eV to -2.5 eV region
%(E$_F$ is taken as the zero of energy), while 
The occupied B $\pi$ states are mostly in the -4 eV to -2 eV region
(E$_F$ is taken as the zero of energy), while $\sigma$ states are
spread throughout the valence band region.  Due to strong
hybridization of the Zr $3d$ states with the B $2p$ states, the
$\sigma$-$\pi$ distinction is not as clear in ZrB$_2$ as it is in
MgB$_2$.  The Zr $3d$ DOS shows that considerably less than 40\% of
the $3d$ DOS is occupied (which would be the case four of ten possible
$3d$ electrons) suggesting Zr$\rightarrow$B charge transfer.
%The hybridization and charge transfer results in the pseudogap separating
%primarily B $2p$ bands (occupied) from primarily Zr $3d$ bands
%(unoccupied).

%FIG. 1
\begin{figure}[tbp]
\epsfxsize=3.1in\centerline{\epsffile{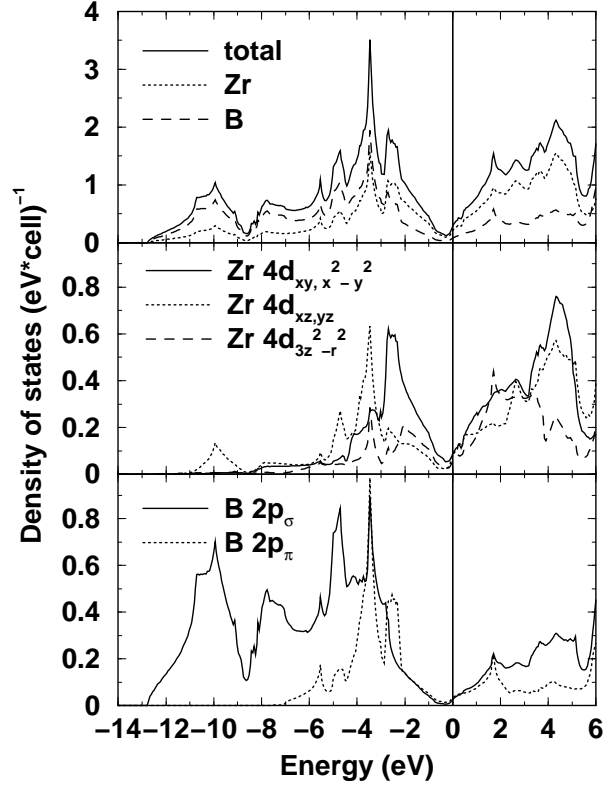}}
%\end{minipage}
%\vspace{3mm}
\caption{
Total and projected density of states for ZrB$_2$.  The projections 
show that there is no distinctive character of states at E$_F$,
rather they are a combination of Zr $4d$ and B $\pi$ and $\sigma$.
}
\label{zrb2dos}
\end{figure}

\begin{figure}[tbp]
\epsfxsize=3.1in\centerline{\epsffile{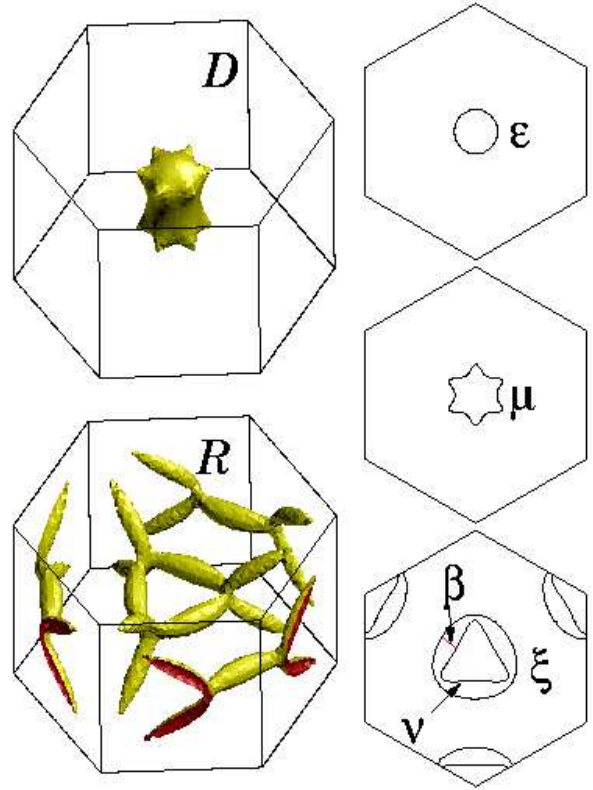}}
\caption{
Calculated Fermi surfaces (left) and selected cross sections (right), 
for ZrB$_2$.  The labels are provided as used in the text.
}
\label{zrb2FS}
\end{figure}

The Fermi surfaces, shown in Fig.\ \ref{zrb2FS}, consist of a
K-centered barbed ring ${\cal R}$ with threefold symmetry, and an
A-centered dumbbell ${\cal D}$ with sixfold symmetry.  We use the
orbit designations of Tanaka and
coworkers.\cite{Zr_tanaka,Zr_ishizawa} For the field along (0001), the
${\cal R}$ surface gives a rounded triangular orbit ($\nu$) encircled
by a nearly circular orbit ($\xi$), while the ${\cal D}$ surface gives
a circular orbit ($\varepsilon$) around its waist and a smoothed-star
orbit ($\mu$) at each end.  These cross sections are also shown in
Fig.\ \ref{zrb2FS}.  In addition, we consider the $\beta$ orbit for
field along ($10\bar 10$), which is the cross section of the barbed
ring ${\cal D}$ in Fig.\ \ref{zrb2FS}.  These Fermi surfaces, which
are broadly consistent with those used by Tanaka and coworkers to
interpret their dHvA data.  The Fermi surfaces presented by Shein and
Ivanovskii (Fig. 1 in Ref.~\cite{shein}) are quite different.  The
origin of this difference is unclear, since the full potential linear
muffin-tin orbital method they use should give the same results as our
methods.  We note that their value N(E$_F$) = 0.163 eV$^{-1}$ differs
considerably from our results and most of the previous calculations.
\cite{ihara,harmon,vajeeston}

%FIG. 2

The general shape of the density of states, and the low value of
N(E$_F$), was confirmed by xray photoemission spectroscopy
measurements.\cite{ihara} The most noteworthy feature, in light of a
new report\cite{gasparov} of T$_c$ = 5.5 K, is the small calculated
value of N(E$_F$) = 0.26 /eV-cell, corresponding to a bare linear
specific heat coefficient $\gamma_b$ = 0.61 mJ/mole-K$^2$.  [The value
of N(E$_F$) is somewhat sensitive to the quality of the calculation.]
The reported experimental\cite{tyan} value of $\gamma$ = 0.47
mJ/mole-K$^2$, which should include electron-phonon enhancement, is
{\it smaller} than our bare band structure value.

The low value of N(E$_F$) itself and the weak
coupling\cite{helgefuture} seem inconsistent with T$_c$ = 5.5 K; it
would require very large electron-phonon matrix elements which has
been shown not to be the case in another transition metal diboride
TaB$_2$.\cite{tab2}
%In addition, the reported superconductivity was found to be sensitive to 
%unexpectedly small magnetic fields.  
Since all information is consistent with the observed
superconductivity arising from a minority phase in the sample, and the
calculations seem to make make superconductivity unlikely, we conclude
that ZrB$_2$ itself is not superconducting.  These facts, and results
presented below, suggest that remeasurement of the heat capacity on
additional samples may be called for.  In any case, the occurrence (or
not) of superconductivity in ZrB$_2$ is peripheral to the intent of
this paper.

%FIG. 3
\begin{figure}[tbp]
\epsfxsize=3.1in\centerline{\epsffile{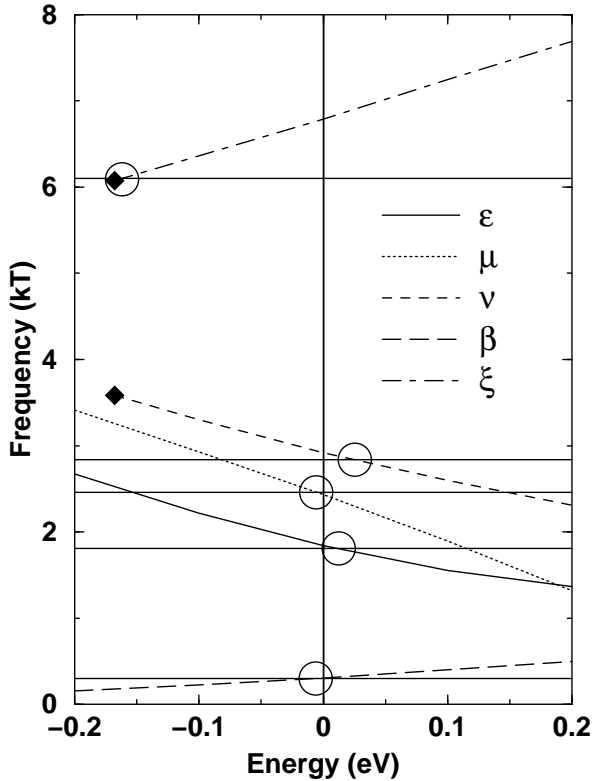}}
%\psfig{figure=dFdE_zrb2.eps,width=6.5cm}
%\end{minipage}
%\vspace{3mm}
\caption{
The areas F(E) of selected extremal constant energy surfaces in ZrB$_2$
for field along (0001), versus energy (E$_F$ = 0).  Sloping lines show
our calculated values, experimental numbers are denoted by horizontal
lines, and the circles denote the energy where perfect agreement occurs.
The three smaller are in good agreement with experiment; the discrepancy
for the large surface is discussed in the text.
}
\label{ZrdFdE}
\end{figure}

The calculated and observed areas are given in Table I.  Percentage
discrepancy or even absolute differences do not give the most physical
indication of the level of (dis)agreement, especially when areas get
small (see MgB$_2$, below).  To indicate more clearly the implication
of the discrepancy between theory and observation, in Fig.\
\ref{ZrdFdE} the orbit areas F(E) versus energy are provided.  The
observed areas are shown as horizontal lines, which allows one to read
off the shift in any given band to bring its area into agreement with
the observed area. The band masses, proportional to the derivative
$dF/dE$, are also readily obtained from such curves.

For the orbits $\varepsilon, \mu, \nu$ in the range 1.8 kT $<$ F $<$ 3
kT, the agreement is excellent, probably within the total numerical
precision.  For the $\xi$ orbit, the agreement is not good: the
reported area is 10\% smaller than we calculate (this amounts to 1.4\%
of the BZ basal plane area).  Moreover, it is unlikely that even this
relatively modest disagreement can be achieved by a shift of the band,
because such a shift destroys agreement for the $\nu$ orbit and
results in a change of topology of the constant energy surfaces.  To
investigate this discrepancy further, we have calculated the $\beta$
orbit around the ${\cal R}$ surface, which passes through both the
$\xi$ and $\nu$ orbits.  This area is in excellent agreement with
experiment (see Table I).  It may be possible, but seems unlikely,
that this disagreement could be reconciled by some non-rigid shift(s)
of band(s).  It is possible that this discrepancy is due to
experimental conditions: the $\xi$ orbit is the largest, giving the
fewest oscillations to fit to an oscillatory form, hence larger
uncertainty in the result.  An experimental reinvestigation of the
dHvA frequencies is under way \cite {winzer02} to either confirm or
resolve this discrepancy.

\subsection{MgB$_2$}
To facilitate understanding of notation, we identify the orbits by
their B character ($\sigma$ or $\pi$) and by the point in the BZ
around which they are centered.  The $\sigma$ Fermi surfaces, pictured
in several previous publications,\cite{kortus,yildirim} are two
concentric fluted cylinders oriented along $\Gamma$-A, which give rise
to extremal orbits $\sigma_{\Gamma}^{S}$, $\sigma_{\Gamma}^{L}$,
$\sigma_{A}^{S}$, $\sigma_{A}^{L}$ for magnetic field along
(0001). ($S, L$ denotes small, large.)  The $\pi$ bands give rise to
$\pi_{\Gamma}$ and $\pi_{A}$ for field along (0001), $\pi_M$ for field
along (1${\bar 1}$00), and $\pi_{L}$ for field along (1000).

The calculated areas and band masses are given in Table II, with
comparison to the three orbits of Yelland {\it et al.}\cite{dhva}
assuming the same correspondence of observed orbit and calculated
orbit.  For reference, an area of 1 kT corresponds to 2\% of the area
A$_{BZ}$ of the basal plane of the BZ.  For the three observed orbits,
the calculated areas are 0.30$\pm$0.04 kT larger than observed, {\it
i.e.} a discrepancy equal to 0.6\% of A$_{BZ}$.  Our calculated areas
are in good agreement (usually close to significant digits) with those
of Mazin and Kortus.\cite{mazin} There are differences compared to the
areas presented by Elgazzar {\it et al.}\cite{elgazzar} (most of our
areas are $\sim$20\% smaller), presumably due to the approximations
made in their augmented spherical wave method.

\begin{figure}[tbp]
\epsfxsize=3.1in\centerline{\epsffile{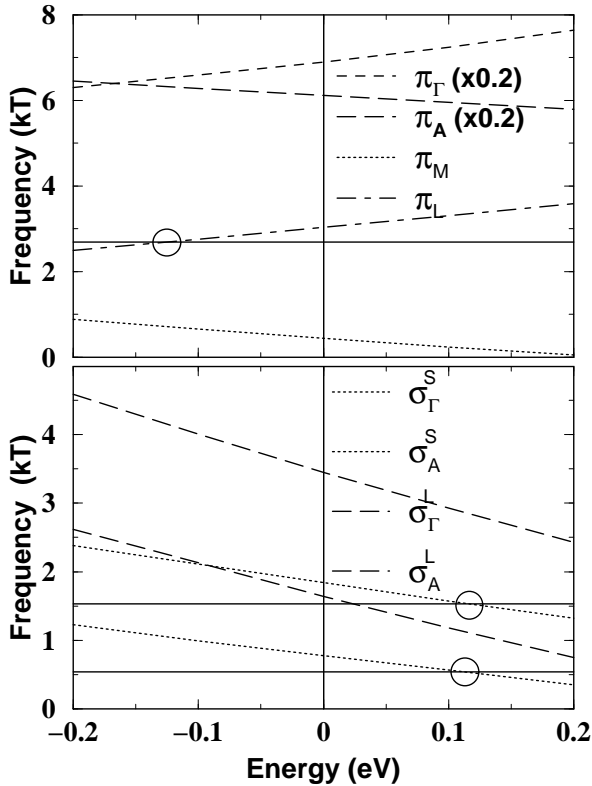}}
%\psfig{figure=dFdE_mgb2.eps,width=6.5cm}
%\end{minipage}
%\vspace{3mm}
\caption{
The areas F(E) of selected extremal constant energy surfaces as in
Fig.\ \ref{ZrdFdE}, but for MgB$_2$.  Bands shifts of about -115 meV
for the $\sigma$ bands and +125 meV for the $\pi$ bands bring the
calculated areas into perfect agreement with the data.
}
\label{MgdFdE}
\end{figure}

In Fig.~\ref{MgdFdE} the orbit areas F(E) versus energy are provided.  The band
masses $m_b = (\hbar^2/2\pi)dF/dE$ are seen to be insensitive to the
position of the Fermi energy except for the two largest $\pi$ orbits.
These values are in good, but not quite perfect, agreement with those
of Mazin and Kortus.  The observed areas (Table II) are shown as
horizontal lines, which allows one to read off the shift in any given
band to bring its area into agreement with the observed area.  (Due to
the difference in the $\sigma$ and $\pi$ densities of states at E$_F$,
charge balance requires a readjustment of E$_F$ by $\sim$20 meV.)  The
required energy shift is -115 meV for both $\sigma_{\Gamma}^{S}$ and
$\sigma_A^{S}$, and +125 meV for $\pi_L$, both of which reduce the
sizes of the hole and electron Fermi surfaces.  These required shifts
imply: (1) there is a relative shift of the $\sigma$ bands relative to
the $\pi$ bands by $\sim$ 240 meV presumably due to correlation
effects beyond LDA, and (2) the $k_z$ dispersion of the $\sigma$ bands
along the $\Gamma$-A line is described correctly in band theory (the
same $\sigma$ band shift is required at $\Gamma$ and at A).  The data
give the areas for only a single tube (the smaller one), but since the
two $\sigma$ are degenerate along $\Gamma$-A, both bands must shift
together.
%The dHvA data therefore imply that there is a relative
%shift of the $\sigma$ and $\pi$ bands ($\sigma$ bands downward, 
%$\pi$ bands upward, all
%Fermi surfaces getting smaller) that represents an exchange-correlation
%correction beyond LDA.

%FIG. 4

The observed effective masses include an enhancement due (primarily)
to electron-phonon coupling, which is given in Table II and is
obtained from $m^* = m_b (1+\lambda$).  For both $\sigma$ orbits the
enhancement thus derived is $\lambda_{\sigma}$ =1.3; for the $\pi$
orbit it is $\lambda_{\pi}$=0.5.  These $\sigma$ orbit enhancements
are noticeably larger than the average values over each surface $\bar
\lambda_{\sigma} = 0.9\pm 0.1, \bar \lambda_{\pi} = 0.4\pm 0.1$
obtained from solution to the anisotropic Eliashberg equations with
first principles band theory input by Choi {\it et al.}\cite{choi} The
difference particularly for the $\sigma$ bands might indicate there is
some feature of strong coupling in MgB$_2$ yet to be understood, or
possibly that the $\sigma$ band shift necessary to provide agreement
with orbit areas, when taken into account in the calculations of
electron-phonon coupling, will ameliorate this discrepancy.  The
disagreement is small enough that the general picture -- the
quasiparticle band structure and electron-phonon coupling determined
within conventional (LDA) band theory, then T$_c$ determined by
Eliashberg theory -- seems well justified.

\section{Discussion}
For ZrB$_2$ the only discrepancy between calculated and observed areas
occurs for the largest orbit that was observed, the overall agreement
between the calculated and the observed frequencies is of the same
order as the achievable numerical accuracy.  Because of the strong
hybridization between the Zr $3d$ and the B $2p$ states, ZrB$_2$ has a
much more isotropic electronic structure than does MgB$_2$.  As a
result, any renormalization and shift of bands with respect to each
other seems to be negligible for this compound, and the LDA single
particle picture description is well justified.

While the observed dHvA areas of MgB$_2$ are readily understood in
terms of the calculated Fermi surfaces and the inferred
electron-phonon coupling constants are reasonable, complete agreement
of the FS areas with the data requires a shift of the $\sigma$ bands
by 115 meV downward, and a shift of the $\pi$ bands upward by 125 meV.
The necessary shift results in $\sigma$ band edges at $\Gamma$ and A
that are reduced 0.38 eV $\rightarrow$ 0.25 eV, 0.76 eV $\rightarrow$
0.63 eV, respectively.  The volume of the $\sigma^S$ tube,
proportional to the average of the two areas in Table II, then is
about 30\% smaller than given by the band calculation, so the number
of holes decreases proportionally (from 0.146 to 0.106).

We can suggest at least two possible causes of this ``beyond LDA''
correction to the band structure.  One possibility is related to the
observation that the charge in the $\sigma$ bands is confined to the
two dimensional B sheet of graphene structure, nearly filling those
bonding states.  Being of more localized character than the $\pi$
bands, the exchange potential in LDA may be less accurate than for the
$\pi$ electrons; a better exchange potential would be larger in
magnitude (and attractive), lowering the $\sigma$ bands with respect
to the $\pi$ bands.  A related viewpoint of the same physics is that
there is a larger (spurious) self-interaction for the $\sigma$ states
than for the $\pi$ states in LDA.

Another possible correction could arise from residual hole-hole
Coulomb interactions in the $\sigma$-hole gas.  There is some analogy
with a related situation in metallic, ferromagnetic Ni, where there
are $\sim$0.6 $3d$ holes/Ni in the minority bands.  Unlike in MgB$_2$,
in Ni there is presumably a relatively large value of the ratio $U/W$
(intra-atomic Coulomb interaction U, $3d$ bandwidth W) making
Hubbard-type correlations of some relevance.  MgB$_2$ has broad
$\sigma$ bands, so Hubbard-like correlations should not be the
problem.  The strong 2D character of the $\sigma$ band holes may
enhance many body corrections.  It will require further work to
determine whether it is one of these mechanisms, or perhaps some
other, that is responsible for the band structure corrections.

\section{Summary}
The rather close correspondence of the LDA Fermi surfaces of MgB$_2$
supports the prevalent picture of superconductivity that is based on
the LDA starting point: very strong electron-phonon coupling of
$\sigma$ band hole states to certain phonons (the E$_{2g}$ branch).
Of the suggested alternatives cited in the Introduction, the
comparison we have made tends to rule out several of them.  Although
the $\sigma$ band Fermi energy requires some adjustment to account
for the observed dHvA orbital areas, it remains around 0.5 eV, large
enough to render non-adiabatic processes unimportant in
superconducting pairing.  The half-filled $\pi$ ideas and resonating
valence bond state idea certainly get no support from the
correspondence of dHvA data to the band theory results.  Other
experimental data (penetration depth, specific heat, isotope shift,
superconducting gap and T$_c$ itself) are being shown in many studies
to require an anisotropic (two band) model, and that most data seem to
be consistent with such models based in detail on band theory results.
Single crystal data, of which the dHvA data\cite{dhva} are some of the
first, will serve to clarify these issues further.

%\section{Summary}
\section{Acknowledgment}
We thank A.\ Carrington, J.R.\ Cooper, H.\ Eschrig, I.\ Mazin, K.\
Winzer, H.\ Uchiyama, P.M.\ Oppeneer, and S.V.\ Shulga for helpful discussions.
This work was supported by National Science Foundation Grant
DMR-0114818 (W.E.P.), and by the Deutscher Akademischer
Austauschdienst (H.R.).

%\vskip -5mm
% References

%\end{multicols}

\begin{table}
\caption{Calculated de Haas-van Alphen areas and masses of ZrB$_2$, compared to
the experimental data of Tanaka and Ishizawa.
Areas are quoted in kTesla.}
\vspace{2mm}
\begin{tabular}{c|llr}
Orbit & F$_{exp}$ & F$_{calc}$ & $m_b$  \\
\tableline
$\varepsilon$(0001) & 1.81~ & 1.84~ & -0.38~  \\
$\mu$(0001)         & 2.46~ & 2.43~ & -0.60~  \\
$\nu$(0001)         & 2.84~ & 2.92~ & -0.41~  \\
$\xi$(0001)         & 6.09~ & 6.78~ & ~0.51~  \\
$\beta(10{\bar 1}0)$& 0.300 & 0.305 & ~0.103
\end{tabular}
\end{table}

\begin{table}
\caption{Calculated de Haas-van Alphen parameters of MgB$_2$ compared to the
experimental data.  
Areas are quoted in kT.  Column 1: the calculated FPLO areas, using the LDA
potential calculated using 16221 k points in the irreducible BZ; values in 
parentheses are from FLAPW using the GGA exchange-correlation.  
Column 2 and column 4: data from Yelland
{\it et al.}  Column 3: band mass, in units of electron mass (negativ
masses stand for holes). Column
5: orbit-averaged mass enhancement $\lambda = |m^*|/|m_b| - 1$.}
\vspace{2mm}
\begin{tabular}{c|lrrrr}
Orbit              & F$_{calc}$ & F$_{exp}$ & $m_b$ & $|m^*|$ & $\lambda$ \\
\tableline
$\sigma_{\Gamma}^{S}$& 0.78 (0.79) & 0.54  &-0.25 & 0.57 & 1.3  \\
$\sigma_{\Gamma}^{L}$& 1.65 (1.67) &       &-0.57 &      &      \\
$\pi_{\Gamma}         $&34.5~ &       & 1.87 &      &      \\
$\sigma_{A}^{S}     $& 1.83 (1.81) & 1.53  &-0.31 & 0.70 & 1.3  \\
$\sigma_{A}^{L}     $& 3.45 (3.46) &       &-0.64 &      &      \\
$\pi_{A     }         $&30.6~ &       &-0.93 &      &      \\
$\pi_{M     }         $& 0.45 &       &-0.25 &      &      \\
$\pi_{L     }         $& 3.03 & 2.69  & 0.32 & 0.47 & 0.5  \\
\end{tabular}
\end{table}

\end{document}